\documentclass[conference]{IEEEtran}

\ifCLASSINFOpdf

\else

\fi

\usepackage{graphicx}

\usepackage{amssymb,amsfonts, amsmath, amsthm}
\usepackage{soul}
\usepackage{xcolor}
\usepackage{hyperref}

\newtheorem{lemma}{Lemma} 
\usepackage{algorithm,algpseudocode}
\algrenewcommand\algorithmicindent{0.7em}
\usepackage{cite}
\usepackage[labelsep=space, font=small, labelfont=bf]{caption} 
\usepackage{subcaption}

\begin{document}
%
\title{Statistical Linear Regression Approach to Kalman Filtering and Smoothing under Cyber-Attacks
}


\author{
\IEEEauthorblockN{Kundan Kumar, Muhammad Iqbal, and Simo Särkkä}
\IEEEauthorblockA{Department of Electrical Engineering and Automation, Aalto University, Finland} 
\{kundan.kumar, muhammad.iqbal, simo.sarkka\}@aalto.fi
}

\maketitle

\begin{abstract}
Remote state estimation in cyber-physical systems is often vulnerable to cyber-attacks due to wireless connections between sensors and computing units. In such scenarios, adversaries compromise the system by injecting false data or blocking measurement transmissions via denial-of-service attacks, distorting sensor readings. This paper develops a Kalman filter and Rauch--Tung--Striebel (RTS) smoother for linear stochastic state-space models subject to cyber-attacked measurements. We approximate the faulty measurement model via generalized statistical linear regression (GSLR). The GSLR-based approximated measurement model is then used to develop a Kalman filter and RTS smoother for the problem. The effectiveness of the proposed algorithms under cyber-attacks is demonstrated through a simulated aircraft tracking experiment. 
\end{abstract}


\IEEEpeerreviewmaketitle

\section{Introduction}
\label{sec:intro}
Cyber-physical systems (CPSs) are integral to applications such as intelligent transportation, terrestrial exploration, power grids, aerospace, and hazardous environments \cite{hu2016estimation, zhang2015survey, wu2018secure, khalid2016bayesian, zhang2012network}. These systems rely on wireless communication for remote monitoring and control, making them vulnerable to cyber-attacks. Adversaries can manipulate sensor measurements by injecting false data or blocking transmissions, thereby compromising system performance. In the absence of such adversarial disruptions, state estimation methods such as the Kalman filter (KF) and Rauch–Tung–Striebel (RTS) smoother provide optimal solutions for linear Gaussian systems \cite{rauch1965maximum, sarkka2023bayesian}. However, their performance degrades significantly under cyber-attacks, necessitating robust estimation techniques capable of handling compromised measurements.


Several types of cyber attack models have been studied in the literature, including replay attacks \cite{mo2009secure}, denial-of-service (DoS) attacks \cite{zhang2015optimal}, and false data injection (FDI) attacks \cite{chen2022class, singh2021bayesian}. In a replay attack, the intruder observes and records measurement readings for a certain time period and later retransmits the recorded measurements to carry out the attacks \cite{mo2009secure}. The DoS attacks \cite{zhang2015optimal} block measurement transmission from the sensor to the computing unit. In FDI attacks, the intruders have access to the measurements and modify them in a random manner \cite{mo2010false, chen2022class}. In this manuscript, we focus on the state estimation under the DoS and FDI attacks.


A few works have addressed the estimation problem under DoS and FDI attacks. The authors in \cite{zhang2015optimal} proposed an optimal attack strategy for an energy-constrained attacker in a linear state space model (SSM). In \cite{li2015jamming}, a game-theoretic approach is used to develop a state estimation algorithm for linear SSM under DoS attacks. For the linear system dealing with the additive FDI attacks, the state estimation algorithms have been developed for sensor networks \cite{mo2010false} and power grids \cite{liu2011false}. A risk-sensitive filtering algorithm for an inaccurate linear system model under additive FDI attacks was proposed in \cite{kumar2024risk}. To handle additive FDI attacks in nonlinear state-space models, estimation algorithms based on the extended Kalman filter (EKF), unscented Kalman filter (UKF), and cubature Kalman filter (CKF) have been developed in \cite{liu2016extended, lu2021unscented, lv2022adaptive}. Recently, \cite{chen2022class, singh2021bayesian} proposed a generalized Bayesian filtering framework to address simultaneous additive and multiplicative FDI attacks. A state estimation algorithm under DoS and additive FDI attacks was developed in \cite{kumar2024gaussian}. However, there is a lack of estimation methods that account for DoS and simultaneous additive and multiplicative FDI attacks. This paper aims to address this gap.

In this paper, we develop a Kalman type of filter and RTS smoother for state space models under DoS and FDI attacks. First, we approximate the faulty measurement model \cite{chen2022class} using the generalized statistical linear regression (GSLR) approach. Subsequently, the GSLR-based approximated measurement is used to develop the estimation algorithm under the Bayesian framework. The main contributions of this article are as follows: (1) We present a unified formulation of the faulty measurement model that accounts for DoS and simultaneous additive and multiplicative FDI attacks. 
(2) Using the GSLR method, we approximate the faulty measurement model.
(3) Based on the approximated measurement model, we develop the Kalman filter and RTS smoother.
(4) Finally, we demonstrate the performance of the proposed algorithm through numerical experiments.


 \begin{figure*}
		\centering
		\setlength{\unitlength}{0.23in} 
		\begin{picture}(25, 18) 
        \put(2,12.5){Process}
		\put(0, 10.75){\framebox(6, 1.5){$x_{k}=A_{k-1} x_{k-1} +\eta_{k-1}$}}
            \put(7.75, 11.5){\rotatebox{150}{\rlap{\makebox[0.3cm]{\hrulefill}}}}
        \put(7.75, 11.5){\rotatebox{180}{\rlap{\makebox[1cm]{\hrulefill}}}}
         \put(7.75, 11.5){\rotatebox{210}{\rlap{\makebox[0.3cm]{\hrulefill}}}}
        \put(6.5,11.75){$x_k$}
  
        \put(9,12.5){Sensor}
		\put(7.75, 10.75){\framebox(4.2, 1.5){$z_k=H_k x_k+\nu_k$}}
        \put(13.65, 11.5){\rotatebox{150}{\rlap{\makebox[0.3cm]{\hrulefill}}}}
        \put(13.65, 11.5){\rotatebox{180}{\rlap{\makebox[1cm]{\hrulefill}}}}
         \put(13.65, 11.5){\rotatebox{210}{\rlap{\makebox[0.3cm]{\hrulefill}}}}
         \put(12.5,11.75){$z_k$}
         
        \put(13.65, 10.75){\rotatebox{90}{\rlap{\makebox[0.25cm]{\hrulefill}}}}
        \put(13.65, 11.275){\rotatebox{90}{\rlap{\makebox[0.25cm]{\hrulefill}}}}
        \put(13.65, 11.8){\rotatebox{90}{\rlap{\makebox[0.25cm]{\hrulefill}}}}
        \put(14.08, 10.75){\rotatebox{180}{\rlap{\makebox[0.25cm]{\hrulefill}}}}
        \put(14.608, 10.75){\rotatebox{180}{\rlap{\makebox[0.25cm]{\hrulefill}}}}
        \put(15.136, 10.75){\rotatebox{180}{\rlap{\makebox[0.25cm]{\hrulefill}}}}
        \put(15.664, 10.75){\rotatebox{180}{\rlap{\makebox[0.25cm]{\hrulefill}}}}
        \put(16.192, 10.75){\rotatebox{180}{\rlap{\makebox[0.25cm]{\hrulefill}}}}
        
        \put(14.08, 12.25){\rotatebox{180}{\rlap{\makebox[0.25cm]{\hrulefill}}}}
        \put(14.608, 12.25){\rotatebox{180}{\rlap{\makebox[0.25cm]{\hrulefill}}}}
        \put(15.136, 12.25){\rotatebox{180}{\rlap{\makebox[0.25cm]{\hrulefill}}}}
        \put(15.664, 12.25){\rotatebox{180}{\rlap{\makebox[0.25cm]{\hrulefill}}}}
        \put(16.192, 12.25){\rotatebox{180}{\rlap{\makebox[0.25cm]{\hrulefill}}}}
        \put(16.17, 10.75){\rotatebox{90}{\rlap{\makebox[0.25cm]{\hrulefill}}}}
        \put(16.17, 11.275){\rotatebox{90}{\rlap{\makebox[0.25cm]{\hrulefill}}}}
        \put(16.17, 11.8){\rotatebox{90}{\rlap{\makebox[0.25cm]{\hrulefill}}}}
        \put(13.9,11.6){{Wireless}}
        \put(14.05,11){{channel}}
        \put(14.9, 12.25){\rotatebox{60}{\rlap{\makebox[0.3cm]{\hrulefill}}}}
        \put(14.9, 12.25){\rotatebox{90}{\rlap{\makebox[1.13cm]{\hrulefill}}}}
         \put(14.9, 12.25){\rotatebox{120}{\rlap{\makebox[0.3cm]{\hrulefill}}}}
        \put(12.70,14.5){DoS \& FDI attacks }

        \put(17.9, 11.5){\rotatebox{150}{\rlap{\makebox[0.3cm]{\hrulefill}}}}
        \put(17.9, 11.5){\rotatebox{180}{\rlap{\makebox[1cm]{\hrulefill}}}}
         \put(17.9, 11.5){\rotatebox{210}{\rlap{\makebox[0.3cm]{\hrulefill}}}}
         \put(16.75,11.75){$y_k$}
        \put(17.9, 10.5){\framebox(2, 2){}}
        \put(18.1,11.65){GSLR}
        \put(18,11.15){approx.}

         \put(22.98, 13.175){\rotatebox{186.5}{\rlap{\makebox[0.3cm]{\hrulefill}}}}
        \put(22.98, 13.175){\rotatebox{208.5}{\rlap{\makebox[2.05cm]{\hrulefill}}}}
         \put(22.98, 13.175){\rotatebox{230.5}{\rlap{\makebox[0.3cm]{\hrulefill}}}}
         \put(22.98, 10.335){\rotatebox{137}{\rlap{\makebox[0.3cm]{\hrulefill}}}}
        \put(22.98, 10.335){\rotatebox{159}{\rlap{\makebox[1.92cm]{\hrulefill}}}}
         \put(22.98, 10.335){\rotatebox{181}{\rlap{\makebox[0.3cm]{\hrulefill}}}}

        \put(23.25,14.2){Filtering}
        \put(23.05,12.45){\framebox(3.25,1.5){$p(x_k\mid y_{1:k})$}}
        \put(23.25,11.35){Smoothing}
        \put(23.05,9.6){\framebox(3.25,1.5){$p(x_k\mid y_{1:T})$}}

		\end{picture}	
		\vspace{-5.55cm}
		\caption{Schematic diagram of the proposed estimation algorithm under DoS and FDI attacks. The filtering (forward pass) and smoothing algorithms are developed based on the GSLR-based approximated faulty measurement model.}
		\label{schematic_diag_FDIA}
	\end{figure*}
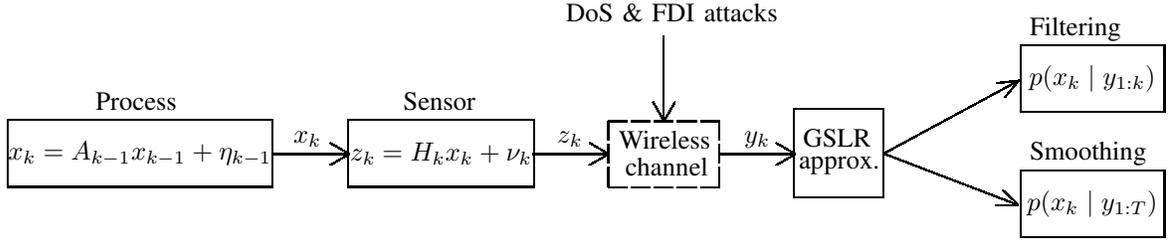

\section{Problem Formulation}\label{sec_prob_form}
Consider a stochastic dynamical system with the following state space model \cite{bar2004estimation,sarkka2023bayesian}
	\begin{align}
	x_{k} & = A_{k-1} x_{k-1} + \eta_{k-1}, \label{Process_eqn}\\
	z_k  & = H_k x_k + \nu_k, \label{measurement_eqn}
	\end{align}
	where $x_k\in \mathbb{R}^{n_x}$, and $z_k\in \mathbb{R}^{n_z}$ are the state of the dynamic system and sensor measurement, respectively. Here, the matrix $A_{k-1} \in \mathbb{R}^{n_x \times n_x}$ is the state transition matrix of the dynamic model, $H_k \in \mathbb{R}^{n_z \times n_x}$ is the measurement model matrix. The terms $\eta_{k-1} \sim \mathcal{N}(0, \, Q_{k-1})$ and $\nu_k \sim \mathcal{N}(0, \,R_k)$ represent the Gaussian process noise and measurement noise, respectively. The initial state $x_0 \sim \mathcal{N}(\hat{x}_{0 \mid 0}, \, P_{0 \mid 0})$, and the noises $\eta_{k-1}$ and $\nu_k$ are mutually independent.

\subsection{Measurement model under DoS and FDI attacks}
Sensor measurements transmitted over a wireless communication channel to a remote computing unit are vulnerable to adversarial manipulation \cite{mo2009secure, zhang2015optimal, chen2022class, singh2021bayesian} (see Fig. \ref{schematic_diag_FDIA}). Attackers can alter the measurements through DoS and FDI attacks \cite{chen2022class, singh2021bayesian, kumar2024gaussian}. 
We model the faulty measurement under DoS and FDI attacks as follows:
\begin{equation}
\begin{split}
    y_k & =  \xi_{b, k} z_k + (1 - \xi_{b, k}) \xi_{c, k} \Big\{ \xi_{m, k} m_k (z_k + \xi_{a, k} a_k) \\& 
    \quad + (1 - \xi_{m, k}) (z_k + \xi_{a, k} a_k) \Big\}, 
\end{split} 
\label{measurement_FDIA0}
\end{equation}
where $y_k \in \mathbb{R}^{n_z}$ is the measurement after the attack, $a_k \sim \mathcal{N}(\mu_a, \Sigma_a)$ and $m_k \sim \mathcal{N}(\mu_m, \sigma_m^2)$ are additive and multiplicative false data parameter to alter the sensor measurement. The terms $\xi_{a, k}$, $\xi_{b, k}$, $\xi_{c,k}$ and $\xi_{m, k}$ in the model are Bernoulli random variables (BRVs) with probabilities $\alpha_a$, $\alpha_b$, $\alpha_c$, and $\alpha_m$, respectively.
The model given in \eqref{measurement_FDIA0} can further be simplified to
 \begin{equation}\label{measurement_FDIA}
     \begin{split}
        y_k & = \xi_{b, k} z_k + (1 - \xi_{b, k}) \xi_{c, k}  \left(1 + \xi_{m, k} \left(m_k - 1\right) \right) \\& \times (z_k + \xi_{a, k} a_k).
     \end{split}
 \end{equation}
 Table \ref{tab_attack} presents attack scenarios under various stochastic parameter values. For special cases of the attacks model in \eqref{measurement_FDIA}, see the references in the rightmost column of Table \ref{tab_attack}. The faulty measurement model given in \eqref{measurement_FDIA} combines the models proposed in \cite{chen2022class} and \cite{kumar2024gaussian} into a single model. 





\begin{table}[]
\caption{Various attack types under different stochastic parameters}
    \centering
    \begin{tabular}{|c|c|c|}
    \hline
         \multicolumn{2}{|c|}{Parameters} & Attack types \\
         \hline
        \multicolumn{2}{|c|}{$\xi_{b, k} =1$,\,\, $\xi_{c, k}, \xi_{a,k}, \xi_{m, k} \in \{0, 1\}$} & No attack \\
         \hline
          &  $\xi_{a, k}=1, \xi_{m, k}=0$ & Additive FDIA \cite{singh2021bayesian, kumar2024gaussian} \\
         \cline{2-3}
         $\xi_{b, k} =0, \xi_{c, k} =1$&$\xi_{a, k}=0, \xi_{m,k}=1$&Multiplicative FDIA \cite{singh2021bayesian} \\\cline{2-3}
         &$\xi_{a, k}=1$, $\xi_{m, k}=1$& Simultaneous FDIA \cite{chen2022class}\\\hline
         \multicolumn{2}{|c|}{$\xi_{b, k} =0, \xi_{c, k} =0$, \, $\xi_{a,k}, \xi_{m, k} \in \{0, 1\}$} & DoS attack \cite{kumar2024gaussian} \\
         \hline
    \end{tabular}
    \label{tab_attack}
\end{table}

\section{Approximating the Faulty Measurement using GSLR Method}\label{sec3_FD_meas_appr}
In this section, we aim to approximate the faulty measurement model \eqref{measurement_FDIA} using GSLR \cite{sarkka2023bayesian,tronarp2018iterative, hostettler2020importance} with respect to the probability density function (pdf) $\mathcal{N}(x_k\mid \hat{x}_{k\mid k-1}, \, P_{k\mid k-1})$ as follows:
\begin{equation}\label{GSLR_meas_GMM}
y_k \approx H_k^{+} x_k + b_k^{+} + \tilde{\nu}_k,
\end{equation}
where $H_k^{+} \in \mathbb{R}^{n_z \times n_x}$ and $b_k^{+} \in \mathbb{R}^{n_z}$ are the coefficients of the approximation, and $\tilde{\nu}_k \sim \mathcal{N}(0, \tilde{\Omega}_k)$ represents the error, consisting of approximation error and measurement noise. The associated parameters of the approximated measurement model \eqref{GSLR_meas_GMM} can be expressed as 
\begin{equation}\label{parameter_FDIA_GSLR}
\begin{split}
    H_k^{+} & = P_{k\mid k-1}^{yx} P_{k\mid k-1}^{-1}, \\
    b_k^{+} & = \hat{y}_{k\mid k-1} - H_k^{+} \hat{x}_{k\mid k-1}, \\
    \tilde{\Omega}_k & = P_{k\mid k-1}^{yy} - H_k^{+} P_{k\mid k-1} H_k^{+\top}. 
\end{split}
\end{equation}
In \eqref{parameter_FDIA_GSLR}, the associated moments can be computed as follows:
\begin{align}
     & \hat{y}_{k\mid k-1}  = E_{\text{pr}}\left[  E\left[y_k \mid x_k\right] \right] \label{m_pi_y},\\&
  P_{k\mid k-1}^{yy}  = E_{\text{pr}}\left[ \mathbb{V}\left[y_k \mid x_k\right]    \right] +
  \mathbb{V}_{\text{pr}}\left[E \left[y_k\mid x_k\right] \right], \label{Pyy_k}\\&
   P_{k\mid k-1}^{yx}   = \mathbb{C}_{\text{pr}}\left[E\left[y_k\mid x_k\right], \, x_k\right] \label{Pyx_pi},
\end{align}
where $E_{\text{pr}}\left[ \cdot \right]$, $\mathbb{V}_{\text{pr}}\left[ \cdot \right]$, $\mathbb{C}_{\text{pr}}\left[ \cdot \right]$ denote expectation, variance, and covariance, respectively, with respect to the prior pdf $ \mathcal{N}(x_k\mid \hat{x}_{k\mid k-1},\, P_{k\mid k-1})$. Before computing the moments mentioned above, we introduce the following useful lemma.
\begin{lemma}\label{Lem1}
Consider the probability distribution of attack parameters, $a_k \sim \mathcal{N}(\mu_a, \Sigma_a)$, $m_k \sim \mathcal{N}(\mu_m, \sigma_m^2)$ and BRVs $\xi_{a,k}, \, \xi_{b, k}, \, \xi_{c, k}, \, \xi_{m,k}$. Then, we have the following: 
    \begin{equation} \label{lem_1a}
    \begin{split}
         & E\left[\left( 1-\xi_{b, k} \right)^2 \xi_{c, k}^2 \big( 1+ \xi_{m, k} (m_k-1) \big)^2\right] \\& = (1-\alpha_b) \alpha_c \left[ (1-\alpha_m) + \alpha_m (\sigma_m^2 + \mu_m^2)\right],
    \end{split}
    \end{equation}
    \begin{equation}\label{lem_1b}
    \begin{split}
        & E\Big[\big\{ (1-\xi_{b, k}) \xi_{c, k} (1+ \xi_{m, k} (m_k - 1)) - (1-\alpha_b) \alpha_c \\& \times (1 + \alpha_m (\mu_M -1)) \big\}^2\Big] = 
        (1-\alpha_b) \alpha_c \Big[ \big( (1-\alpha_m) \\& + \alpha_m   (\mu_m^2 + \sigma_m^2) \big) - (1-\alpha_b) \alpha_c \big(1+ \alpha_m (\mu_m -1) \big)^2  \Big],
    \end{split}  
    \end{equation}
    and 
\begin{equation}\label{lem_1c}
\begin{split}
   & E\big[ \big(\xi_{a, k} (a_k - \mu_a) + (\xi_{a, k} - \alpha_a) \mu_a  \big) \big(\xi_{a, k}(a_k - \mu_a)  \\& + (\xi_{a, k} - \alpha_a) \mu_a  \big)^\top\big]   = \alpha_a \Sigma_a + \alpha_a (1-\alpha_a) \mu_a \mu_a^\top. 
\end{split}
\end{equation}    
    \begin{proof}
    First, recall that if $\xi$ is a Bernoulli random variable with probability $\alpha$, we have
     \begin{equation}\label{prob_Ber}
     \begin{split}
         & p(\xi = 1) = E[\xi] = E[\xi^r] = \alpha, \\
       & p(\xi = 0) = E[1 - \xi] = E[(1 - \xi)^r] = 1 - \alpha, \\
      &  E[(\xi - \alpha)^2] = (1-\alpha)\alpha, 
     \end{split}
 \end{equation}
 where $r \geq 2$ is an arbitrary constant. The expected value of the square of $(1-\xi_{b, k}) \xi_{c, k}( 1+ \xi_{m, k} (m_k-1) \big)$ is 
   \begin{equation*}
   \begin{split}
       & E\big[(1-\xi_{b, k})^2 \xi_{c, k}^2 \big( 1+ \xi_{m, k} (m_k-1) \big)^2\big]  =  E\big[ (1-\xi_{b, k})^2 \big] \\& \times E\big[ \xi_{c, k}^2 \big]  E\big[ 1 + \xi_{m, k}^2 (m_k^2 - 2 m_k + 1)  + 2 \xi_{m, k} (m_k -1)  \big].
   \end{split}
    \end{equation*}  
    Using \eqref{prob_Ber} and probability distribution of $m_k$, the above equation becomes \eqref{lem_1a}.
    Now, we calculate 
    \begin{equation*}
        \begin{split}
           & 
           E\Big[\big\{ (1-\xi_{b, k}) \xi_{c, k} \big(1+ \xi_{m, k} (m_k - 1)\big) - (1-\alpha_b) \alpha_c \big(1 +  \alpha_m \\& \times (\mu_M -1)\big) \big\}^2\Big] =  E\Big[(1-\xi_{b, k})^2 \xi_{c, k}^2 \big(1+ \xi_{m, k} (m_k - 1)\big)^2\Big] \\& - 2 E\Big[ (1-\xi_{b, k}) \xi_{c, k} \big(1+ \xi_{m, k} (m_k - 1)\big) \Big] (1-\alpha_b) \alpha_c \big(1 +  \alpha_m \\& \times (\mu_M -1)\big) + (1-\alpha_b)^2 \alpha_c^2 \big(1 +  \alpha_m  (\mu_M -1)\big)^2.      
        \end{split}
    \end{equation*}
    Utilizing \eqref{prob_Ber} along with the distribution of $m_k$, the above equation simplifies to \eqref{lem_1b}. We compute the expected value of the given below equation as 
    \begin{equation*}
    \begin{split}
        & E\big[ \big(\xi_{a, k} (a_k - \mu_a) + (\xi_{a, k} - \alpha_a) \mu_a  \big) \big(\xi_{a, k}(a_k - \mu_a) \\& +  (\xi_{a, k} - \alpha_a) \mu_a  \big)^\top\big]   
        = E\big[ \xi_{a, k}^2 (a_k - \mu_a)(a_k - \mu_a)^\top  \big]  \\& +
         E\big[ \xi_{a, k} (\xi_{a, k} - \alpha_a)  (a_k - \mu_a)\mu_a^\top  \big]  + E\big[ \xi_{a, k}  (\xi_{a, k} - \alpha_a) \\& \times \mu_a (a_k - \mu_a)^\top  \big] + E\big[(\xi_{a, k} - \alpha_a)^2 \mu_a \mu_a^\top  \big].
    \end{split}
\end{equation*}
Following the properties of $\xi_{a,k}$ and $a_k$, we receive \eqref{lem_1c}. 
    \end{proof}
\end{lemma}
\subsection{Conditional moment computation}
Next, we compute the moments given in \eqref{m_pi_y}-\eqref{Pyx_pi}, which we use in \eqref{parameter_FDIA_GSLR}. The conditional expectation of $y_k$ given $x_k$ using \eqref{measurement_eqn} and \eqref{measurement_FDIA} is as follows:
\begin{equation}\label{ykmidxk}
\begin{split}
   & E[y_k \mid x_k] 
     = \alpha_b H_k x_k + (1-\alpha_b) \alpha_c \\& \times \left(1 +  \alpha_m (\mu_m -1) \right)  (H_k x_k + \alpha_a \mu_a).
\end{split}
\end{equation}
The expectation of \eqref{ykmidxk} with respect to the pdf $ \mathcal{N}(\hat{x}_{k\mid k-1},\, P_{k\mid k-1})$ is given by
\begin{equation}\label{E_pi_E_y_giv_x}
\begin{split}
   & \hat{y}_{k\mid k-1}  =E_{\text{pr}}[E[y_k \mid x_k]] =  \alpha_b H_k \hat{x}_{k\mid k-1} + (1-\alpha_b) \alpha_c \\& \times \left(1 +  \alpha_m (\mu_m -1) \right)   (H_k \hat{x}_{k\mid k-1} + \alpha_a \mu_a). 
\end{split}
\end{equation}
Now, we aim to compute $\mathbb{V}_{\text{pr}}$. To this end, we calculate the difference between  $E[y_k\mid x_k]$ and $\hat{y}_{k\mid k-1}$, which is given by
\begin{equation}\label{Exp_y_giv_x_dif_Exp_y_nl}
\begin{split}
    & E[y_k\mid x_k] - \hat{y}_{k\mid k-1} = \alpha_b H_k (x_k - \hat{x}_{k\mid k-1}) +(1-\alpha_b) \\& \times \alpha_c  \left(1 +  \alpha_m (\mu_m -1) \right) H_k(x_k - \hat{x}_{k\mid k-1}). 
\end{split}
\end{equation}
Using \eqref{Exp_y_giv_x_dif_Exp_y_nl}, we calculate $\mathbb{V}_{\text{pr}}\left[E\left[y_k\mid x_k\right]\right]$ as follows:
\begin{equation}\label{var_E_y_giv_x_nl}
    \begin{split}
      &  \mathbb{V}_{\text{pr}}\left[E\left[y_k\mid x_k\right]\right]  = 
        E_{\text{pr}}\big[ (E[y_k\mid x_k] - \hat{y}_{k\mid k-1})  (E[y_k\mid x_k]  \\& - \hat{y}_{k\mid k-1})^\top \big]  = 
        \Big(\alpha_b^2 + 2 \alpha_b  (1-\alpha_b)  \alpha_c \big(1 +\alpha_m  (\mu_m -1) \big) \\&  +
 (1-\alpha_b)^2 \alpha_c^2 \big(1   +  \alpha_m (\mu_m -1) \big)^2 \Big)H_k P_{k\mid k-1} H_k^\top. 
    \end{split}
\end{equation}
Next, we aim to compute the variance of $(y_k \mid x_k)$, denoted as $\mathbb{V}[y_k \mid x_k]$. Before doing so, we first determine the following:
\begin{equation*}
\begin{split}
   & y_k - E\left[y_k\mid x_k\right] 
    =(\xi_{b, k} - \alpha_b)H_kx_k + \xi_{b, k} \nu_k + (1-\xi_{b, k}) \xi_{c, k}\\&   \big( 1 + \xi_{m, k} (m_k -1) \big) \Big(\xi_{a, k} (a_k     - \mu_a)  +  (\xi_{a, k} - \alpha_a ) \mu_a + \nu_k \Big) + \\& \Big( (1-\xi_{b, k}) \xi_{c, k} \big(1+ \xi_{m, k} (m_k  -1) \big) - (1-\alpha_b) \alpha_c \big(1+ \alpha_m \\&
    \times (\mu_m - 1) \big)  \Big) (H_k x_k + \alpha_a \mu_a). 
\end{split}
\end{equation*}
Using \eqref{prob_Ber} and Lemma \ref{Lem1}, we compute $\mathbb{V}[y_k \mid x_k]$ as follows: 
\begin{equation*}
    \begin{split}
     &   \mathbb{V}[y_k \mid x_k]  
        = E\left[\left(y_k - E\left[y_k\mid x_k\right]\right)\left(y_k - E\left[y_k\mid x_k\right]\right)^\top \mid x_k\right] \\&
        = E\big[ (\xi_{b, k} - \alpha_b)^2 \big] E\big[ H_k x_k x_k^\top H_k^\top \big] + E[\xi_{b, k}^2] E[\nu_k\nu_k^\top] + \\&  E\left[ (1- \xi_{b, k})^2 \xi_{c, k}^2 \big( 1+ \xi_{m, k} (m_k -1) \big)^2\right] E\Big[ \big( \xi_{a, k} \left(a_k - \mu_a\right)    \\&    +  \left(\xi_{a, k}    - \alpha_a \right) \mu_a \big)  \big( \xi_{a, k} \left(a_k - \mu_a\right) + \left(\xi_{a, k} - \alpha_a \right) \mu_a \big)^\top + \nu_k \nu_k^\top \Big]    \\& + E\left[ \Big( (1-\xi_{b, k}) \xi_{c, k} \left( 1+ \xi_{m, k} (m_k -1) \right) - (1-\alpha_b) \alpha_c (1+ \alpha_m \right. \\& \left. \times (\mu_m - 1))   \Big)^2 \right]  E\left[  (H_k x_k  + \alpha_a \mu_a) (H_k x_k + \alpha_a \mu_a)^\top \mid x_k \right] \\& 
        = (1-\alpha_b) \alpha_b H_kE[x_kx_k^\top]H_k^\top +\alpha_b R_k + (1-\alpha_b) \alpha_c \big( \alpha_m (\sigma_m^2 + \\& \mu_m^2) + (1-\alpha_m) \big) \big(\alpha_a \Sigma_a + \alpha_a  (1-\alpha_a)   \mu_a \mu_a^\top  + R_k\big)  + (1-\alpha_b) \\&  \alpha_c \Big( \big\{ (1-\alpha_m) + \alpha_m  (\mu_m^2 + \sigma_m^2) \big\}  - (1-\alpha_b) \alpha_c (1+ \alpha_m  (\mu_m -\\& 1))^2 \Big)    \big( H_k x_k x_k^\top H_k^\top  + \alpha_a H_k x_k \mu_a^\top   
     + \alpha_a \mu_a x_k^\top H_k^\top + \alpha_a^2 \mu_a \mu_a^\top \big).
    \end{split}
\end{equation*}
The expectation of $\mathbb{V}\left[y_k \mid x_k\right]$ with respect to prior pdf $\mathcal{N}(\hat{x}_{k\mid k-1}, \, P_{k\mid k-1})$ is given below: 
\begin{equation}\label{E_var_y_giv_x_nl}
    \begin{split}
      & E_{\text{pr}}\left[\mathbb{V}\left[y_k \mid x_k\right]\right] = (1-\alpha_b)\alpha_b H_k(P_{k\mid k-1} + \hat{x}_{k\mid k-1}\hat{x}_{k\mid k-1}^\top) \\& 
      \times H_k^\top  + \alpha_b R_k + (1-\alpha_b) \alpha_c \big( \alpha_m (\sigma_m^2 + \mu_m^2) + (1-\alpha_m) \big)  \\&
      \big(\alpha_a \Sigma_a  + \alpha_a  (1-\alpha_a)  \mu_a \mu_a^\top  + R_k\big)  + (1-\alpha_b) \alpha_c \Big( \big\{ (1-\alpha_m) \\&
      + \alpha_m  (\mu_m^2   + \sigma_m^2) \big\} - (1-\alpha_b) \alpha_c (1+ \alpha_m (\mu_m -1))^2  \Big) \\&
      \Big( H_k \big(P_{k\mid k-1} + \hat{x}_{k\mid k-1}  \hat{x}_{k\mid k-1}^\top \big)  H_k^\top  + \alpha_a H_k \hat{x}_{k\mid k-1}\mu_a^\top  \\&
     + \alpha_a \mu_a \hat{x}_{k\mid k-1}^\top H_k^\top + \alpha_a^2 \mu_a \mu_a^\top \Big). 
    \end{split}
\end{equation}
The expression for $ P^{yy}_{k\mid k-1}$ is derived by adding \eqref{var_E_y_giv_x_nl} and \eqref{E_var_y_giv_x_nl} as given in \eqref{Pyy_k}. Using \eqref{Pyx_pi} and \eqref{Exp_y_giv_x_dif_Exp_y_nl}, we compute $P^{yx}_{k\mid k-1}$ as
\begin{equation}\label{Pyx_pi_final_linear}
    \begin{split}
       & P^{yx}_{k\mid k-1} =  E_{\text{pr}}\left[ \left(E[y_k \mid x_k] - \hat{y}_{k\mid k-1} \right) \left(x_k -  \hat{x}_{k\mid k-1} \right)^\top \right]\\&
        = \Big( \alpha_b + (1-\alpha_b) \alpha_c \big(1+  \alpha_m (\mu_m -1) \big) \Big) H_k P_{k\mid k-1}.
    \end{split}
\end{equation}
After computing these moments, we can evaluate the parameters of the GSLR approximated measurement model given in \eqref{parameter_FDIA_GSLR}. Let us denote the set of parameters related to process dynamics, the measurement model, DoS and FDI attacks as $\Theta_k = [A_k, H_k, Q_k, R_k, \alpha_a, \alpha_b, \alpha_c, \alpha_m, \mu_a, \Sigma_a, \mu_m, \sigma_m^2]$. The pseudo-code for computing the GSLR parameters is presented in Algorithm \ref{Algo_linear_GSLR}. 
  \begin{algorithm}[h!] 
    \caption{Computation of GSLR parameters}\label{Algo_linear_GSLR}
		\begin{algorithmic}[1]
  \Function{$[H_k^{+}, b_k^{+}, \tilde{\Omega}_k] = \text{GSLR}$}{$ \hat{x}_{k\mid k-1}, P_{k\mid k-1}, \Theta_k$}.
    \State Obtain $\hat{y}_{k\mid k-1}$, $P^{yy}_{k\mid k-1}$, and $P^{yx}_{k\mid k-1}$ following \eqref{E_pi_E_y_giv_x}, \eqref{Pyy_k}, and \eqref{Pyx_pi_final_linear}, respectively.
    \State Compute $H_k^{+}, b_k^{+},$ and $\tilde{\Omega}_k$ using \eqref{parameter_FDIA_GSLR}. 
  \EndFunction
  \end{algorithmic}
  \end{algorithm}
\section{Kalman Filter and RTS Smoother under Cyber-Attacks}\label{sec4_affine}
In this section, we derive the Kalman filter and RTS smoother based on the linear SSM \eqref{Process_eqn}-\eqref{measurement_eqn} and the GSLR-based approximated faulty measurement model \eqref{GSLR_meas_GMM}. The smoothing algorithm consists of two sequential steps: (i) the forward pass and (ii) the backward pass. In forward pass, the Kalman filtering algorithm is used to compute the prior and posterior moments. 
Using \cite[Lemma A.3]{sarkka2023bayesian}, the marginal distribution of $x_k$ is 
$p(x_{k}\mid y_{1:k-1}) = \mathcal{N}(x_{k}\mid \hat{x}_{k\mid k-1}, P_{k\mid k-1}),$
where 
\begin{align}
     \hat{x}_{k\mid k-1} &= A_{k-1} \hat{x}_{k-1|k-1}  \label{prior_mean_linear}, \\
  P_{k\mid k-1} &= A_{k-1} P_{k-1 \mid k-1} A_{k-1}^\top+ Q_{k-1}\label{prior_cov_linear}.
\end{align}

Next, we compute the joint distribution of the state, $x_k$ and the GSLR-based approximated measurement, $y_k$ given $y_{1:k-1}$ as follows: 
\begin{equation*}
\begin{split}
   & p(x_{k}, y_k\mid y_{1:k-1}) = p(y_k\mid x_{k}) \, p(x_{k}\mid y_{1:k-1}) \\& 
    \approx \mathcal{N}(y_k\mid H_k^{+}x_k + b_k^{+}, \tilde{\Omega}_k) \, \mathcal{N}(x_k\mid \hat{x}_{k\mid k-1}, P_{k\mid k-1}) \\& 
    \approx \mathcal{N}\bigg(\begin{bmatrix}
        x_{k}\\
        y_k
    \end{bmatrix}\bigg|\hat{\mathcal{X}}_{k}, \mathcal{P}_{k} \bigg),
\end{split} 
\end{equation*}
where 
\begin{equation*}\label{prior_cov_forward}
\begin{split}
   \hat{\mathcal{X}}_{k} &= \begin{bmatrix}
        \hat{x}_{k\mid k-1} \\
       H_k^{+} \hat{x}_{k \mid k-1} + b_k^{+}
    \end{bmatrix}, \\
    \mathcal{P}_{k} &= \begin{bmatrix}
        P_{k\mid k-1} & P_{k\mid k-1} H_k^{ + \top}\\
        H_k^{ +} P_{k\mid k-1} & H_k^{ +} P_{k\mid k-1} H_k^{ + \top}+\tilde{\Omega}_k
    \end{bmatrix}.
\end{split}
\end{equation*}
The conditional probability distribution of $x_k$ given $y_{1:k}$ is 
\begin{equation*}\label{post_pdf_filter}
    p(x_k\mid y_{1:k}) \approx \mathcal{N}(x_k\mid \hat{x}_{k\mid k}, P_{k\mid k}), 
\end{equation*}
where 
\begin{align}
    \hat{x}_{k\mid k} &= \hat{x}_{k \mid k-1} + P_{k|k-1} H_{k}^{+\top} (H_{k}^{+}  P_{k|k-1} H_{k}^{+\top} + \tilde{\Omega}_k )^{-1} \nonumber \\& \quad \,\, \times (y_k -  H_{k}^{+} \hat{x}_{k|k-1} - b_{k}^{+}) \label{post_mean_linear},\\
      P_{k\mid k} & = P_{k|k-1} - P_{k|k-1} H_{k}^{+\top} (H_{k}^{+}  P_{k|k-1} H_{k}^{+\top} + \tilde{\Omega}_k )^{-1} \nonumber \\& \quad  \,\, \times H_{k}^{+}  P_{k|k-1} \label{post_cov_linear}.
\end{align} 

After computing $p(x_k \mid y_{1:k}) \approx \mathcal{N}(x_k\mid \hat{x}_{k \mid k}, P_{k \mid k})$ for $k \in \{1, \ldots, T \}$, we perform the backward pass step to recursively compute $p(x_k \mid y_{1:T}) = \mathcal{N}(x_k\mid \hat{x}_{k \mid T}^s, P_{k \mid T}^s)$ starting from $k = T$. Note that the smoothing algorithm is independent of the measurement model, allowing us to directly apply the standard smoothing algorithm for the linear affine SSM \cite[pp. 255–260]{sarkka2023bayesian}. The detailed implementation of the Kalman filtering and smoothing solution for the attacked model is presented in Algorithm \ref{Algo_RTS_linear}. 
 \begin{algorithm}[h!]
		\caption{Kalman filter and RTS smoother for the attacked model}\label{Algo_RTS_linear}
		\begin{algorithmic}[1]
  \Function{$[\hat{x}_{k\mid k}, P_{k\mid k}, \hat{x}_{k\mid T}^s, P_{k \mid T}^s] = \text{KFS}$}{$\hat{x}_{0\mid 0}, P_{0\mid 0}, \Theta_k$}. 
  \For{$k = 1, \ldots, T$}
  \State Compute $\hat{x}_{k\mid k-1}$ and $P_{k\mid k-1}$ using Eqs. \eqref{prior_mean_linear}-\eqref{prior_cov_linear}. 
  \State $[H_k^{+}, \,b_k^{+}, \, \tilde{\Omega}_k] = \text{GSLR}(\hat{x}_{k\mid k-1},  P_{k\mid k-1}, \Theta_k)$.
  \State Compute $\hat{x}_{k\mid k}$ and $P_{k\mid k}$ following Eqs. \eqref{post_mean_linear}-\eqref{post_cov_linear}.
\EndFor
\State $\hat{x}_{T\mid T}^s = \hat{x}_{T\mid T} $ and $P_{T\mid T}^s = P_{T\mid T}$. 
  \For{$k = T-1, \ldots, 1$}
  \State $K_s = P_{k\mid k} A_{k}^\top P_{k+1\mid k}^{-1}$.
  \State $\hat{x}_{k\mid T}^s = \hat{x}_{k\mid k} + K_s (\hat{x}_{k+1\mid T}^s - \hat{x}_{k+1\mid k})$.
  \State $P_{k\mid T}^s  = P_{k\mid k} + K_s (P_{k+1\mid T}^s - P_{k+1\mid k}) K_s^\top$.
\EndFor
\EndFunction
  \end{algorithmic}
  \end{algorithm} 

\section{Simulation Results}\label{sec_sim_res}
   We consider an air-traffic control scenario \cite{arasaratnam2009cubature}, in which an aircraft performs a maneuver in a two-dimensional space with a known turn rate. The discrete-time dynamics of the maneuvering aircraft can be expressed as
\begin{equation*}
    \begin{split}
        x_{k} &= \begin{bmatrix}
            1 & 0 & \frac{\sin \omega t}{\omega} & -\frac{1-\cos \omega t}{\omega} \\
            0 & 1 & \frac{1-\cos \omega t}{\omega} & \frac{\sin \omega t}{\omega}\\
            0 & 0 & \cos \omega t & -\sin \omega t \\
            0 & 0& \sin \omega t & \cos \omega t
        \end{bmatrix} x_{k-1} + \eta_{k-1},
    \end{split}
\end{equation*}
where the state of the aircraft, $x= \begin{bmatrix}
    x_{1}  & x_{2} & \dot{x}_{1} & \dot{x}_{2}
\end{bmatrix}^\top$, $(x_{1}, \, x_{2})$ and $(\dot{x}_{1}, \, \dot{x}_{2})$ represent the position and velocity of the aircraft in the $x$ and $y$ directions, respectively, $t = 0.05$ s is the sampling time, and $\omega = 3^\circ \, \text{/s}$ is turn rate. The process noise $\eta_{k-1} \sim \mathcal{N}(0,  \, Q)$ with $Q = \text{diag}(0.3^2,\, 0.3^2,\, 0.05^2,\, 0.05^2)$. The measurement available to the estimator is the position of the aircraft, that is 
\begin{equation*}
        z_k = \begin{bmatrix}
            I_{2\times 2} & 0_{2 \times 2}
        \end{bmatrix} x_k + \nu_k, 
\end{equation*}
where $\nu_k \sim \mathcal{N}(0, R)$ with $R = \text{diag}(12, \, 12)$. We perform the simulation for 20 s. The following parameters associated with the cyber-attacks are used in the simulation: $\alpha_a = 0.3$, $\alpha_b = 0.7$, $\alpha_c = 0.9$,  $\alpha_m = 0.1$, $m_k \sim \mathcal{N}(0.95, \, 0.10^2)$ and $a_k \sim \mathcal{N}(\mu_a, \Sigma_a)$, where $\mu_a = \begin{bmatrix}
    0.7 &  0.9
\end{bmatrix}^\top$ and $\Sigma_a = \text{diag}(1, \, 0.5)$. We initialize the estimator with $x_0 = \begin{bmatrix}
    200 \, \text{m} & 200 \, \text{m} & 15\, \text{m/s} & 15\, \text{m/s}
\end{bmatrix}^\top$, $\hat{x}_{0\mid 0} = \begin{bmatrix}
    250 \, \text{m} & 150 \, \text{m} & 12\, \text{m/s} & 17\, \text{m/s}
\end{bmatrix}^\top$, and $P_{0 \mid 0} = \text{diag}(10^2 \, \text{m}^2, \, 10^2 \,\text{m}^2,\, 4^2 \, \text{m}^2/\text{s}^2,\, 4^2 \, \text{m}^2/\text{s}^2)$.

    \begin{figure}
    \centering
    \includegraphics[width=8cm, height = 5 cm]{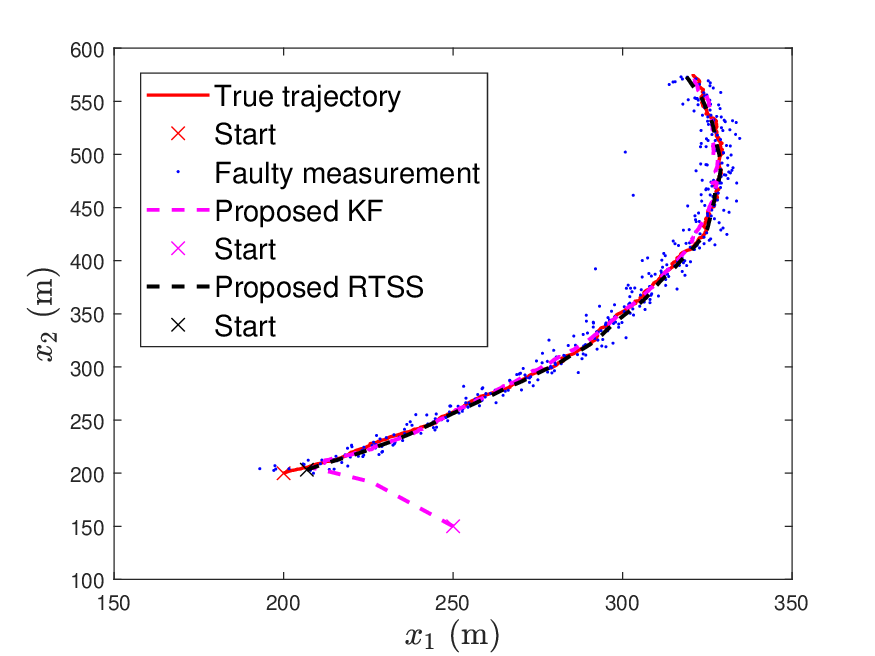}
    \caption{The true trajectory, faulty measurement, the proposed KF and RTSS for the aircraft tracking problem in the presence of cyber-attacks in a single representative run.}
    \label{truth_car_track}
\end{figure}
\begin{figure}[htb]
   \centering 
\begin{minipage}[b]{.49\linewidth}
  \centerline{\includegraphics[width=4.9cm, height = 3.5 cm]{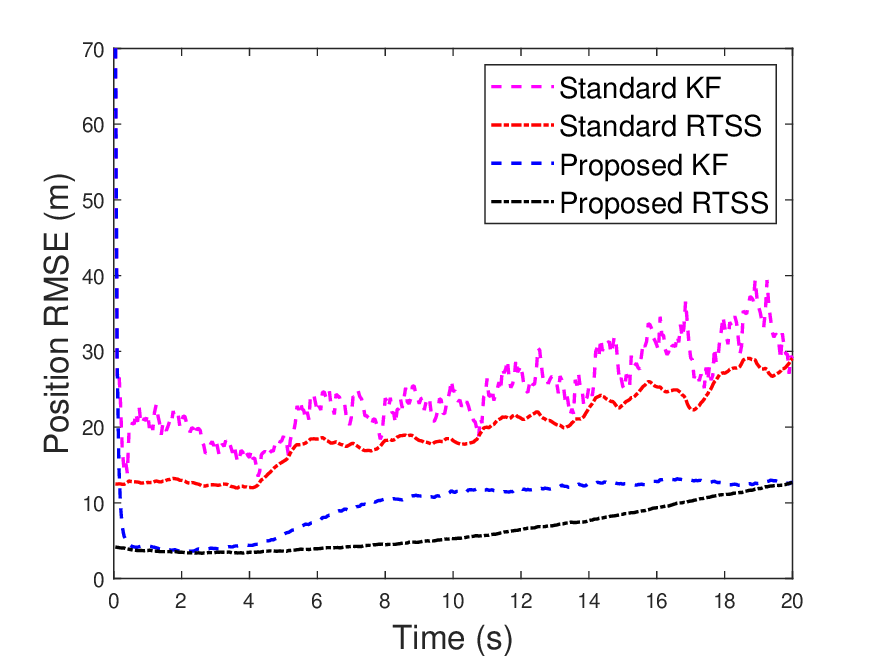}}
  \centerline{(a) Position RMSE}\medskip
\end{minipage}
\begin{minipage}[b]{0.49\linewidth}
\centerline{\includegraphics[width=4.9cm, height = 3.5 cm]{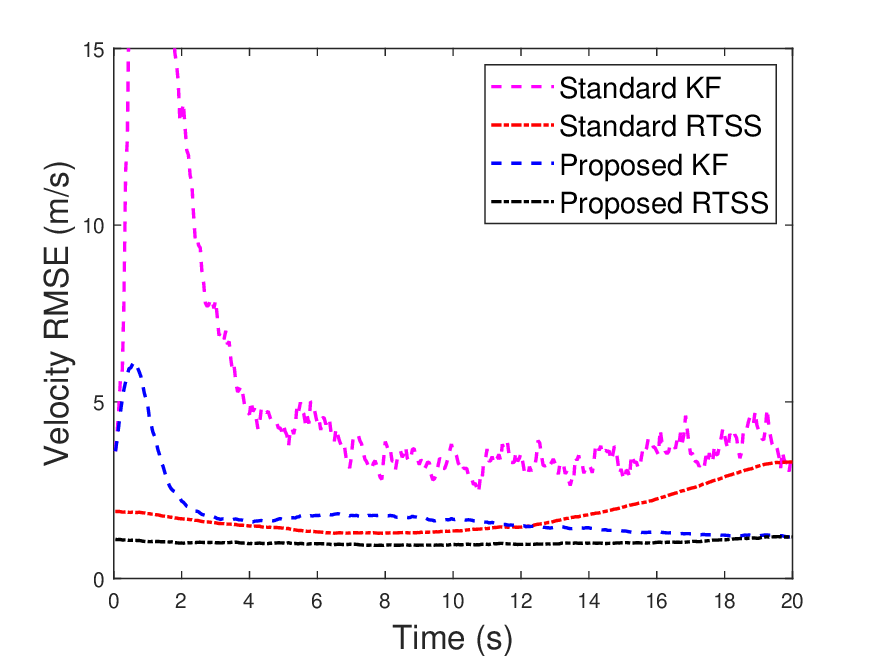}}
  \centerline{(b) Velocity RMSE}\medskip
\end{minipage}
\caption{The position and velocity RMSE of different estimators for the aircraft tracking problem under DoS and FDI attacks, obtained from 100 MC runs.}
\label{RMSE_car_track_simultaneous_attack}
\end{figure} 

We implemented the standard KF, the standard RTSS, the proposed filtering algorithm (that is, the forward pass of Algorithm \ref{Algo_RTS_linear}), and the RTS smoother, referring to them as the proposed KF and proposed RTSS. Fig. \ref{truth_car_track} shows the simulated true aircraft trajectory, the measurements under cyber-attacks, and the estimated trajectory obtained using the proposed KF and RTSS in a single Monte Carlo (MC) run. Notably, Fig. \ref{truth_car_track} demonstrates that the proposed KF and RTSS effectively track the aircraft even in the presence of cyber-attacks. 

We evaluate the estimators' performance in terms of the position and velocity root mean square error (RMSE). Fig. \ref{RMSE_car_track_simultaneous_attack} presents the position and velocity RMSE of the different estimators obtained from 100 MC runs. The results indicate that the proposed KF and RTSS achieve lower RMSE compared to the standard estimators under cyber-attacks. 


\section{Conclusion}\label{sec_conclusion}
In this article, we have developed a Kalman filter and RTS smoother for the linear SSM under DoS and FDI attacks. We reformulated the faulty measurement model that accounts for DoS and simultaneous additive and multiplicative FDI attacks. We approximated the faulty measurement model using the conditional expectation from the GSLR method. The GSLR-based approximated measurement model was used to derive the filtering and smoothing algorithms. The performance of the methods was illustrated in a simulated aircraft tracking experiment. In the future, we plan to extend this method to nonlinear state-space models. 




\bibliographystyle{IEEEtran}
\bibliography{refs}



\end{document}